\documentclass[final,5p,twocolumn,times]{elsarticle}
\usepackage{amsmath,gensymb,tensind,color}
\usepackage{booktabs}
\usepackage{algorithm}
\usepackage{algorithmic}
\usepackage{url}
\usepackage[nice]{units}

\usepackage[T1]{fontenc}
\usepackage[scaled=0.8]{beramono}

\usepackage{scalefnt}

\newcommand\Comment[1]{}

\newcommand{\vecTwo}[2]{\left(\!\!\begin{array}{c} #1\\ #2\end{array}\!\!\right)}
\newcommand{\diff}{\mathrm{d}}

\newcommand{\verbfont}{\fontsize{8}{10}\selectfont}

\newcounter{bla}

\begin{document}
\tensordelimiter{?}

\title{MPPhys -- A many-particle simulation package for computational physics education}

\author[rvt]{Thomas M{\"u}ller}
\ead{Thomas.Mueller@vis.uni-stuttgart.de}

\address[rvt]{
  Visualisierungsinstitut der Universit\"at Stuttgart (VISUS),
  Allmandring 19, 70569 Stuttgart, Germany
}

\begin{abstract}
In a first course to classical mechanics elementary physical processes like elastic two-body collisions, the mass-spring model, or the gravitational two-body problem are discussed in detail. The continuation to many-body systems, however, is defered to graduate courses although the underlying equations of motion are essentially the same and although there is a strong motivation for high-school students in particular because of the use of particle systems in computer games. The missing link between the simple and the more complex problem is a basic introduction to solve the equations of motion numerically which could be illustrated, however, by means of the Euler method. 
The many-particle physics simulation package \emph{MPPhys} offers a platform to experiment with simple particle simulations.
The aim is to give a principle idea how to implement many-particle simulations and how simulation and visualization can be combined for interactive visual explorations.
\end{abstract}

\maketitle

{\bf PROGRAM SUMMARY}

\begin{small}
\noindent
{\em Program Title:} 
   \\
{\em Catalogue identifier:} 
   \\
{\em Licensing provisions:} 
   \\
{\em Programming language:}
   C++, OpenGL, GLSL, OpenCL
   \\
{\em Computer:} 
   Linux and Windows platforms with OpenGL support
   \\
{\em Operating system:} 
   Linux and Windows
   \\
{\em RAM:} 
   XXX GBytes
   \\
   \\
{\em Classification:} 
  XXX
  \\
{\em External routines/libraries:} 
   OpenGL, OpenCL
   \\
{\em Nature of problem:} 
   integrate n-body simulations, mass-spring models
   \\
{\em Solution method:} 
   Numerical integration of n-body-simulations,
   3D-Rendering via OpenGL.
   \\
{\em Running time:} 
   Problem dependent
   \\

\end{small}

\section{Introduction}
Many-particle simulations that determine the motion of individual particles under their mutual interactions play an important role in numerous applications of chemistry, biology, material sciences, physics, and even computer graphics. 
While the simulation and the subsequent visualization of real systems make great demands on software and hardware, the underlying Newtonian dynamics, however, is easily comprehensible and can be discussed already at high school level.

A step-by-step introduction on how to implement a gravitational $N$-body code starting from high-school level can be found in the online book by Hut and Makino~\cite{hut}. Contrary to standard textbooks, Hut and Makino present the introduction in narrative form where three friends discuss the topic and derive the necessary equations and programming codes.
The ``Molecular Workbench'' (MW)~\cite{mw} is a Java-based learning platform for molecular dynamics simulations, see also Tinker~\cite{tinker2008}. Besides many existing simulations, there is a graphical user interface to create new simulations and embed them into curriculum materials.
A huge collection of diverse physics applications, also in the context of many-particle systems, is provided by the ``Open Source Physics'' (OSP) project~\cite{osp}.

The aim of the many-particle physics simulation package \emph{MPPhys} presented in this paper is in between the above mentioned approaches. In contrast to prefabricated closed applets, the user has full access to the complete programming code to obtain a deeper insight how particle simulations are implemented. The object-oriented structure of \emph{MPPhys}' core functionality facilitates the integration of the equations of motion as well as the interactive visualization of the particles' trajectories for different particle models.
Hardware accelerated integration of the equations of motion is realized using the Open Compute Language (OpenCL).
Additionally, the graphical user interface \emph{QtMPPhys} is a platform for script-based modeling when the actual programming code is of minor interest.
The focus of \emph{MPPhys}, however, lies on the reproducibility of the simulations and visualizations and not in the precision of the integrators or the efficiency of the algorithms. Users concerned with the accuracy of the solutions are encouraged to run the models with varying timesteps as appropriate.


The structure of the paper is as follows. In Section~\ref{sec:implementation} the core functionality of \emph{MPPhys} and the currently implemented particle models are discussed. The graphical user interface \emph{QtMPPhys} is presented in Section~\ref{sec:QtMPPhys}. In Section~\ref{sec:examples}, numerous example models demonstrate the usability of \emph{MPPhys}.

\emph{MPPhys} is implemented in C++ and is freely available for Linux and Windows.
The source code and several examples can be downloaded from \url{go.visus.uni-stuttgart.de/mpphys}.

\section{MPPhys}\label{sec:implementation}
The implementation of \emph{MPPhys} is split into two parts. On the one hand, each particle simulation can be compiled as a standalone program for those who are interested in the implementation itself and those who want direct access to the particle data. Additionally, the particle trajectories can be visualized by means of the Open Graphics Library OpenGL~\cite{opengl}.
\emph{QtMPPhys} on the other hand is a graphical user interface that helps the user to concentrate on the exploration of the different particle systems.
In this section, the details of \emph{MPPhys}' core functionality is highlighted. The graphical user interface is described in Section~\ref{sec:QtMPPhys}.

\subsection{MPPhys' core functionality}\label{subsec:mpphysCore}
\begin{figure}[ht]
  \includegraphics[width=0.49\textwidth]{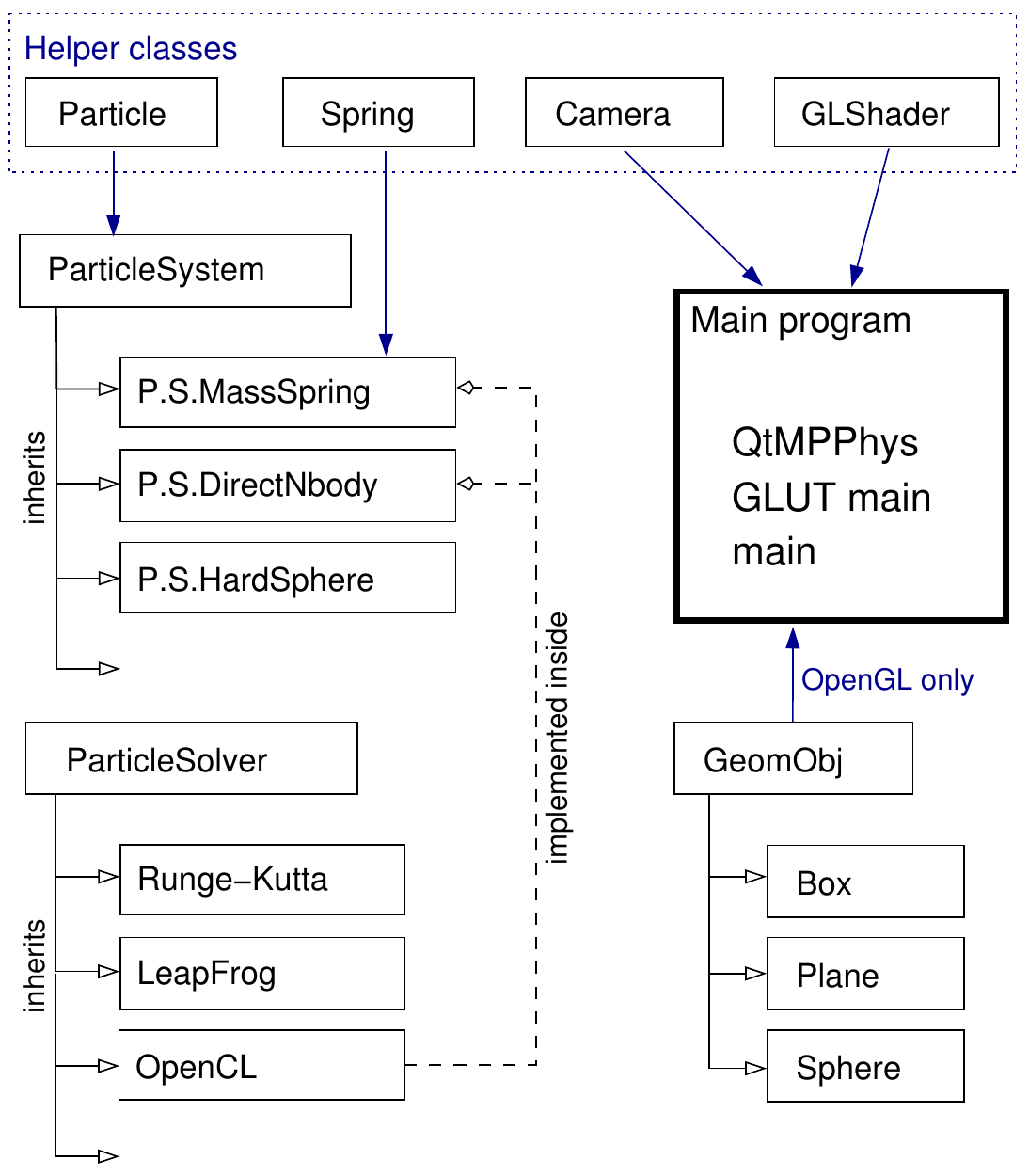}
  \caption{Core structure of \emph{MPPhys}. Any particle system inherits from the base class {\tt ParticleSystem}, and any solver inherits from {\tt ParticleSolver}. However, integrating the particle system by means of OpenCL kernels is implemented within the corresponding particle class. The main program can be either the \emph{QtMPPhys} user interface or a standalone executable for a particular system with or without OpenGL support.}
  \label{fig:mpphysStruct}
\end{figure}

\emph{MPPhys} offers the possibility to explore several different particle systems in one simulation package with full control over the simulation itself and its visualization. For that, the particle systems are implemented following an object-oriented approach, see Fig.~\ref{fig:mpphysStruct}, where each system inherits from the common base class {\tt ParticleSystem}.

Every derived particle class must implement the physics of the particle system it describes and is responsible for rendering the particles on screen using OpenGL if desired. It also has to provide the OpenCL kernels if the integration of the equations of motion can be hardware accelerated. Otherwise, the integration can be hand on to standard CPU solvers like Runge-Kutta or Leap-Frog which themself are implemented in an object-oriented structure with common base class {\tt ParticleSolver}. Particle properties like mass, charge, initial position or velocity, as well as spring properties are stored in the helper classes {\tt Particle} or {\tt Spring}, respectively.

The initial configuration of the particle system can be loaded from plain text files, where each particle is described by the following parameters (see also listing below): initial position {\tt(x,y,z)}, initial velocity {\tt(vx,vy,vz)}, mass {\tt m}, charge {\tt c}, radius {\tt r}, and {\tt color}. The particle motion can be fixed to a specific coordinate direction by setting {\tt 1} for free motion and {\tt 0} if the direction is fixed (currently, this can be used only in {\tt MassSpring} and {\tt DirectNbody} systems). System parameters like the gravitational Gauss constant or the simulation time step can also be defined.
{\verbfont
\begin{verbatim}
# filename: data.plist
#---- ID   x   y   z     m    c   r     color (rgb)
  pos  0  0.0 0.0 0.0   1.0  0.0 0.1    1.0 1.0 0.0 
  pos  1  1.0 0.0 0.0   3e-6 0.0 0.03   0.3 0.3 1.0 
#---- ID  vx   vy     vz
  vel  0  0.0  0.0   0.0
  vel  1  0.0 0.0172 0.0
#---- ID  x y z  // fixed=0, free=1
  vfix 0  1 1 0 
  vfix 1  1 1 0
#---- gravitational Gauss constant
  k2  2.9584e-4
#---- time step  
  dt  0.03
\end{verbatim}
}

Algorithm~\ref{alg:minSetup} shows the minimum setup for an $N$-body simulation of a planet orbiting a star in the $xy$-plane. Line (1) creates an instance of the gravitational $N$-body particle system. The particle data text file is read via line (2). In the setup step, line (3), the particle data is mapped to data arrays that are easier and faster to handle, in particular, when uploading to the graphics board (GPU). The integrator for the particle system, here the Leap-Frog integrator, is chosen in line (4). Now, the system can be integrated step-by-step until a fixed maximum time is reached. For later exploration of the particle trajectories, the positions are stored to file.
\begin{algorithm}[ht]
  \tt
  \begin{algorithmic}[1]
  \verbfont
    \STATE ParticleSystem* ps = new ParticleSystemDirectNbody();    
    \STATE ps->ReadData("data.plist");
    \STATE ps->Setup();
    \STATE ps->InitSolver("LF");
    \WHILE { ps->GetCurrentTime()<50000 }
    \STATE    ps->TimeStep(); 
    \STATE    // output to file
    \ENDWHILE
  \end{algorithmic}
  \caption{Minimum setup.}
  \label{alg:minSetup}
\end{algorithm}

\begin{figure}[ht]
    \includegraphics[width=0.49\textwidth]{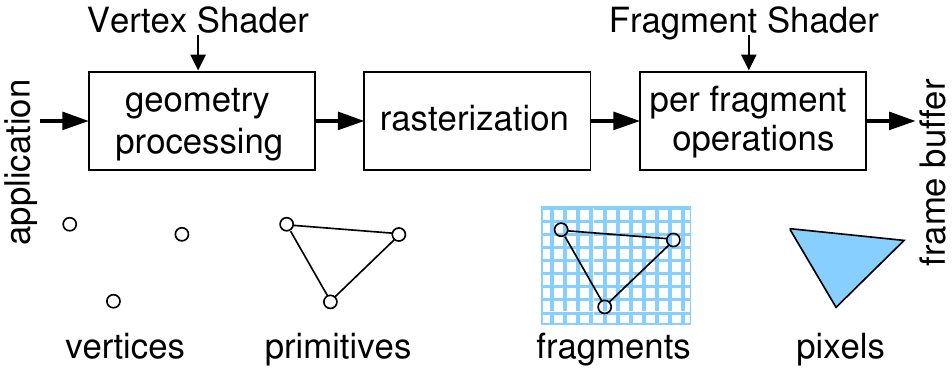}
    \caption{Minimal graphics pipeline. The vertex shader processes the vertices and maps them to window coordinates. After rasterization, the fragment shader can manipulate/colorize the individual fragments.}
    \label{fig:oglpipeline}
\end{figure}

For a direct visual exploration of the particle simulation, an OpenGL~\cite{opengl} rendering environment has to be established. The necessary window management as well as mouse and keyboard handling can be realized using the OpenGL Utility Toolkit (GLUT)~\cite{glut}.
The minimal graphics pipeline to bring a rudimentary particle visualization onto screen is shown in Fig.~\ref{fig:oglpipeline}. After uploading the particle positions (vertices) onto the GPU, the vertex shader is responsible to transform the vertices from world space coordinates to window coordinates. For that, it makes use of the projection matrix that is delivered by the pinhole camera model implemented within the {\tt Camera} helper class. Automatic primitive assembly and rasterization yields pixel-sized fragments that can be modified by the fragment shader whose output is shown on screen. Both shaders are freely programmable with the C-like OpenGL Shading Language (GLSL)~\cite{glsl}. Reading and compiling the shader code is supported by the {\tt GLShader} class.

\subsection{Gravitational N-body simulations}\label{subsec:directnbody}
The \emph{DirectNbody} model simulates the gravitational interaction between $N$ particles. The orbital motion of each individual particle follows from the forces exerted by all other particles. A straightforward implementation has to calculate for each particle $i$ the sum of $N-1$ gravitational interactions $\vec{F}_{ij}$ between particle $i$ and particle $j$,
\begin{equation}
    \vec{F}_i = \sum\limits_{j=1,j\neq i}^N GM_iM_j\frac{\vec{r}_j-\vec{r}_i}{|\vec{r}_j-\vec{r}_i|^3},\quad i=\{1,\ldots,N\},
    \label{eq:gravN}
\end{equation}
where $G$ is Newton's constant, $M_i$ is the mass and $\vec{r}_i$ is the current position of particle $i$.
Due to Newton's third law, however, the computational effort can be reduced to the half. Nonetheless, it is quadratic in $N$.
Professional numerical $N$-body codes reduce the computational effort by sophisticated algorithms, see for example Aarseth~\cite{aarseth2003}. But such methods are out of the scope of this article.

For numerical calculations, Eq.~(\ref{eq:gravN}) is rewritten as
\begin{equation}
  \vec{a}_i = \frac{\diff^2}{\diff t^2}\vec{r}_i = \sum\limits_{j=1,j\neq i}^N k^2m_j\frac{\vec{r}_j-\vec{r}_i}{\left(|\vec{r}_j-\vec{r}_i|^2+\varepsilon^2\right)^{3/2}},
\end{equation}
where $\vec{F}_i=M_i\diff^2\vec{r}_i/\diff t^2$ and $k^2=GM$. Additionally, the masses $M_i=Mm_i$ were replaced by fractions $m_i$ of a `standard mass' $M$. Furthermore, a softening parameter $\varepsilon$ was added to prevent the denominator to diverge. If $M$ equals the solar mass, $M=M_{\text{sol}}$, then $k^2$ is Gauss' gravitational constant, and times and lengths are measured in days and astronomical units (AU).
The OpenCL implementation of the gravitational $N$-body system is based on the \emph{NVidia GPU Computing SDK}~\cite{nvidiasdk}.

\subsection{Mass-spring simulations}\label{subsec:massSpring}
\begin{figure}[ht]
  \centering\includegraphics[width=0.3\textwidth]{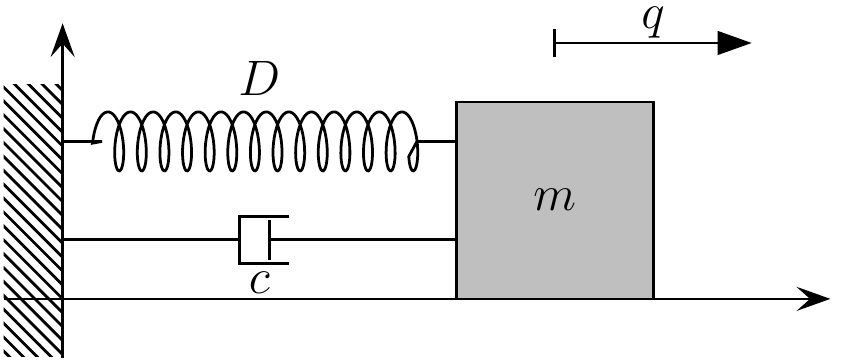}
  \caption{The most simple mass-spring model consists of a mass $m$ that is connected to a wall by a spring with spring constant $D$ and damping factor $c$.}
  \label{fig:massSpringDamp}
\end{figure}

The basis of the \emph{MassSpring} model is the free, damped harmonic oscillator equation. In one dimension this equation reads
\begin{equation}
  m\frac{\diff^2q}{\diff t^2} + c\frac{\diff q}{\diff t} + Dq=0
  \label{eq:fdhOsci}
\end{equation}
with spring constant $D>0$ and velocity-dependent damping factor (frictional coefficient) $c\geq 0$. The coordinate $q$ represents the displacement of the spring from its rest length, see Fig.~\ref{fig:massSpringDamp}.

If the damping factor $c=0$, the solution of Eq.~(\ref{eq:fdhOsci}) is given by $q(t)=a\cos(\omega_0t+\varphi_0)$ with maximum displacement $a$, phase angle $\varphi_0$, angular frequency $\omega_0^2=D/m$, and period $T=2\pi/\omega_0$. If the frictional coefficient is non-vanishing, the solution of (\ref{eq:fdhOsci}) depends on the relation between $\gamma=c/(2m)$ and $\omega_0$. In the weak damping (underdamped) regime, $\gamma^2<\omega_0^2$, the angular frequency reduces to $\omega^2=\omega_0^2-\gamma^2$, and the amplitude decreases exponentially. If the system is overdamped, $\gamma^2>\omega_0^2$, there is no oscillation but an aperiodic creeping. $\gamma^2=\omega_0^2$ is called aperiodic limit case, see e.g. Kuypers~\cite{kuypers}.

In the general situation, there are several masses $m_i$ connected by different springs $(D_{ij},c_{ij})$ for each individual connection between particle $i$ and particle $j$. Then, the equation of motion reads
\begin{equation}
  m_i\frac{\diff^2\vec{x}_i}{\diff t^2}+\sum\limits_{j\in B_i}\left[D_{ij}\frac{\vec{x}_j-\vec{x_i}}{|\vec{x}_j-\vec{r}_i|}\left(|\vec{x}_j-\vec{x}_i|-l_{ij}\right) + c_{ij}\left(\frac{\diff\vec{x}_j}{\diff t}-\frac{\diff\vec{x}_i}{\diff t}\right)\right]=0,
\end{equation}
where $l_{ij}$ is the rest length of the corresponding spring, and $\vec{x}_i$ is the actual position of particle $i$. The sum is over all connections between particle $i$ and particles $j$ being part of the index set $B_i$.

\subsection{Discrete elements with hard spheres}
\begin{figure}[ht]
  \includegraphics[width=0.49\textwidth]{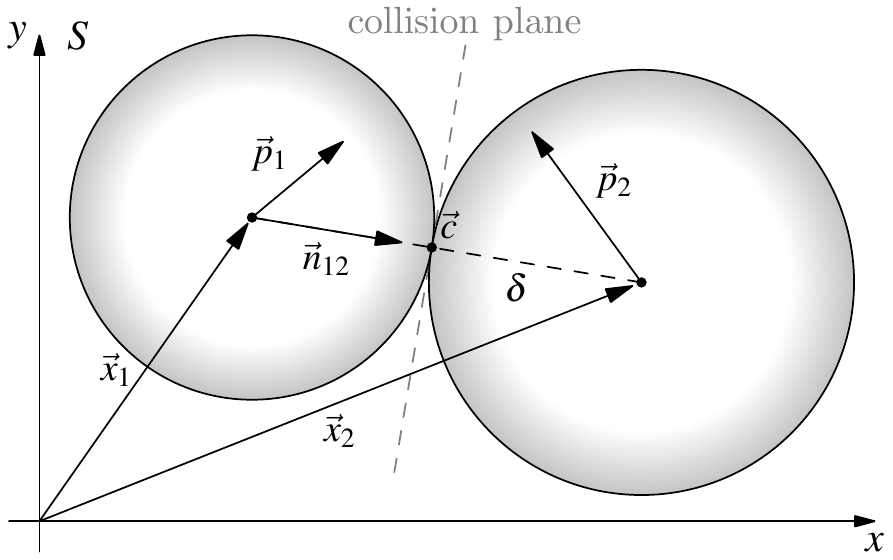}
  \caption{Colliding spheres with masses $m_1$,$m_2$ and radii $r_1$,$r_2$. The position vectors $\vec{x}_1$,$\vec{x}_2$ and the momenta $\vec{p}_1$,$\vec{p}_2$ are given with respect to the laboratory system $S$. The distance between the spheres' centers is denoted by $\delta=r1+r2$, and the point of collision is given by $\vec{c}=\vec{x}_1+r_1(\vec{x}_2-\vec{x}_1)/|\vec{x}_2-\vec{x}_1|=\vec{x}_1+r_1\vec{n}_{12}$.}
  \label{fig:collSpheres}
\end{figure}

A first step to the discrete element method is to use hard spheres of fixed size and mass which move freely until they interact via elastic collisions, see Fig.~\ref{fig:collSpheres}. The interaction happens instantaneously in consideration of energy and momentum conservation,
\begin{equation}
  \vec{p}_1+\vec{p}_2=\vec{p}'_1+\vec{p}'_2,\quad \frac{m_1}{2}v_1^2+\frac{m_2}{2}v_2^2 = \frac{m_1}{2}{v_1}'^2+\frac{m_2}{2}{v_2}'^2,
\end{equation}
with $\vec{p}_i=m_i\vec{v}_i$ being the momentum of particle $i$. Unprimed (primed) coordinates represent velocities and momenta before (after) the collision.

To actually calculate the momenta after the collision, it is necessary to switch to the center-of-mass (CM) system,
\begin{equation}
  \vec{x}_{CM} = \frac{m_1\vec{x}_1+m_2\vec{x}_2}{m_1+m_2},\quad \vec{v}_{CM} = \frac{m_1\vec{v}_1+m_2\vec{v}_2}{m_1+m_2}.
\end{equation}
The velocities of both spheres in the CM system, $\vec{w}_i=\vec{v}_i-\vec{v}_{CM}$, are parallel, $\vec{w}_1||\vec{w}_2$, and the corresponding momenta read $\vec{q}_1=\mu\left(\vec{v}_1-\vec{v}_2\right)$ and $\vec{q}_2=-\mu\left(\vec{v}_1-\vec{v}_2\right)$ with the reduced mass factor $\mu=m_1m_2/(m_1+m_2)$. Then, energy and momentum conservation in the CM system yields
\begin{equation}
  \vec{q}'_1 = -\vec{q}'_2,\quad |\vec{q}_1| = \pm |\vec{q}'_1|,\quad |\vec{q}_2| = \pm |\vec{q}'_2|,
\end{equation}
which fixes the momenta up to a collision angle $\theta$. In the sphere-sphere collision, this angle is naturally fixed requiring that the incident angle equals the emergent angle with respect to the collision plane, see Fig.~\ref{fig:collSpheres}. Thus,
\begin{equation}
  \vec{q}'_1 = \vec{q}_1 - 2\left<\vec{n}_{12},\vec{q}_1\right>\vec{n}_{12},\quad \vec{q}'_2 = \vec{q}_2 - 2\left<\vec{n}_{12},\vec{q}_2\right>\vec{n}_{12}.
\end{equation}
Now, switching back to the laboratory system delivers the velocities after the collision,
\begin{subequations}
\label{eq:velAfter}
\begin{align}
  \vec{v}'_1=\vec{v}_1-2\left<\vec{n}_{12},\frac{m_2\left(\vec{v}_1-\vec{v}_2\right)}{m_1+m_2}\right>\vec{n}_{12},\\
  \vec{v}'_2=\vec{v}_2-2\left<\vec{n}_{12},\frac{m_1\left(\vec{v}_2-\vec{v}_1\right)}{m_1+m_2}\right>\vec{n}_{12},
\end{align}
\end{subequations}
In the special case of equal masses, $m_1=m_2$, zero initial velocity of sphere $2$, $\vec{v}_2=\vec{0}$, and head-on collision, $\vec{n}_{12}=\vec{v}_1/|\vec{v}_1|$, the initially moving sphere transfers its total energy and momentum to the resting sphere: $\vec{v}'_1=\vec{0}$ and $\vec{v}'_2=\vec{v}_1$.

For a continuously progressing simulation, that is inevitable for a smooth visualization, and to be consistent with the other particle systems, the \emph{HardSpheres} model is integrated also in between the collision events. After every time step, all spheres are tested for mutual intersections, $|\vec{x}_i-\vec{x}_j|<r_i+r_j$. If an intersection is detected, the velocities are changed following Eq.~(\ref{eq:velAfter}) and the positions are adapted to correct for the overlap $\delta=r_1+r_2-|\vec{x}_1-\vec{x}_2|$. Hence,
\begin{equation}
  \vec{x}'_n = \vec{x}_n + \Delta t_n\left(-\vec{v}_n + \vec{v}'_n\right), \quad n=\{1,2\},
\end{equation}
with $\Delta t_n = r_n\delta/(r_1+r_2)/\left<\vec{n}_{12},\vec{v}_n\right>$ being the time between the actual contact of the two spheres and the end of the time step due to integration. From the actual contact position $\vec{x}_n-\Delta t_n\vec{v}_n$, the sphere is shifted along the new direction by $\Delta t_n\vec{v}'_n$. 
However, this procedure is numerically unstable because the $\left<\vec{n}_{12},\vec{v}_n\right>$ term in $\Delta t_n$ can become very small. Hence, the position is adapted only approximately by $\vec{x}'_n = \vec{x}_n - r_n\delta/(r_1+r_2)\vec{n}_{12}$.
Furthermore, highly symmetric situations like the first shot in pool billards cannot be simulated exactly because, on the one hand, the \emph{HardSpheres} model takes only two-body interactions/collisions into account, and on the other hand, the mutual particle interaction algorithm is asymmetric by construction.

\section{The graphical user interface QtMPPhys}\label{sec:QtMPPhys}
\begin{figure}[ht]
  \includegraphics[width=0.48\textwidth]{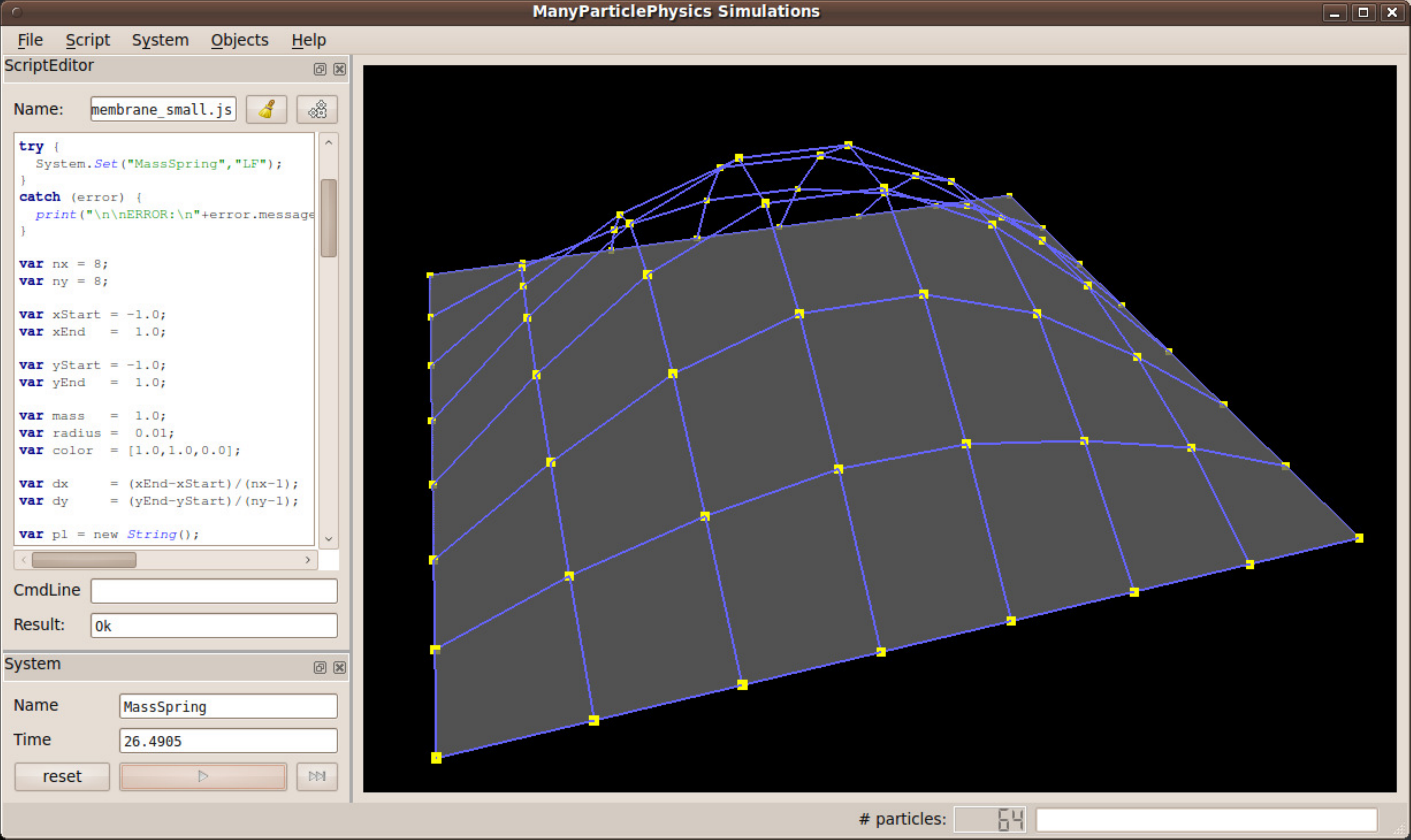}
  \caption{Screenshot of \emph{QtMPPhys} with the main OpenGL window, the script editor, and the system widget with animation control.}
  \label{fig:mpphysGUI}
\end{figure}

The graphical user interface (GUI) of \emph{QtMPPhys} in standard configuration is composed of the OpenGL window, the script editor, and the animation control widget, see Fig.~\ref{fig:mpphysGUI}.

\emph{QtMPPhys} starts in the empty mode. A particle system can be selected, for example, via ``System/Set System'' from the menu bar, where at the same time the integrator has to be chosen. ``System/Load particle data'' from the same menu bar loads the data as described in Section~\ref{subsec:mpphysCore}. Pressing the ``play'' button in the system widget integrates the particle system and shows the animation in the OpenGL window.

\subsection{Script engine}
The particle system selection as well as the definition of the initial conditions and viewing parameters can be all set at once by means of the Qt script engine which is based on the ECMAScript~\cite{ecma} standard. The general procedure, however, is the same. First, the particle system and the integrator has to be chosen, and afterwards, the particle data is handed over as a text string. The script engine can also be used to manipulate system parameters while the simulation is running. For example, the gravitational field or the frictional constant of the springs can be changed to study the influence on the particle system.
An example script is printed in \ref{appsec:scriptExp}.

\subsection{Particle inspector}
\begin{figure}[ht]
  \includegraphics[width=0.49\textwidth]{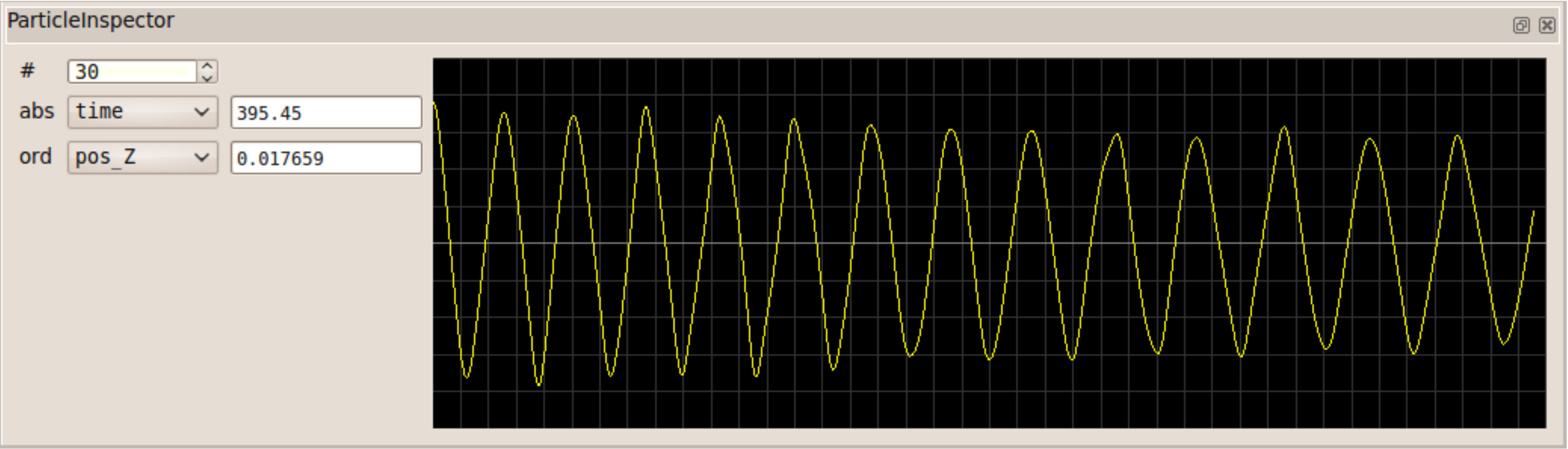}
  \caption{The particle inspector shows particle properties in relation to each other. Here, the $z$-coordinate (ordinate) of particle {\tt \#30} is related to the simulation time (abscissa).}
  \label{fig:partInsp}
\end{figure}

Several particle properties can be investigated visually during the simulation by means of the particle inspector, see Fig.~\ref{fig:partInsp}, where two properties can be shown in relation to each other. The most expedient relation is, for example, a property with respect to the simulation time. In particular, the energy conservation of the complete system can be checked (if implemented in the particle class).

\section{Examples}\label{sec:examples}
In the following, several examples are presented for the currently implemented particle models. The accompanying scripts can be found in the \texttt{jscripts} folder.

\subsection{Sun-Earth system}
The most simple gravitational $N$-body system consists of only two bodies: a massive central star and an orbiting planet of negligible mass. Such a two-body problem can be cast into an effective one-body problem as is done in every book to classical mechanics.
The Sun-Earth system can be realized by means of the \emph{DirectNbody} model where $M=M_{\text{sol}}$, $m_1=1$  (Sun) and $m_2=1/330000$ (Earth). In the simplified situation where the Earth moves on a circular orbit, its initial velocity at distance $r=1$ (AU = astronomical unit) follows from $v^2=k^2/r$ with Gauss' constant $k=0.0172~\textrm{AU}^{3/2}/\text{day}$. Then, the period for one orbit is $T=2\pi/v\approx 365.3$ days. 

Now, you could add the other planets of the solar system and watch the influence of their mutual gravitational interactions. As long as the masses of the planets are used, the solar mass dominates the orbital motion and the gravitational disturbances are observable only in a detailed inspection of the logged data.
The simulation becomes more interesting if some of the masses, for example the mass of Jupiter, will be increased.

\subsection{Circumbinary planetary system}
Until recently, the existence of a planet around a binary star system was only a topic of science fiction. But despite such systems might be counterintuitive, they are realized in nature, see e.g. Doyle et al.~\cite{doyle2011} or Orosz et al.~\cite{orosz2012}.

Because the mass of the planet $m_P$ is much less than the mass of the two stars, $m_A$ and $m_B$, the motion of both can be described as two-body problem which itself can be cast into an effective one-body problem by separating the center of mass motion. With 
\begin{equation}
    \vec{r}_{cm} = \frac{m_A\vec{r}_A+m_B\vec{r}_B}{m_A+m_B},\quad m_A\ddot{\vec{r}}_A=\vec{F}_{BA},\quad m_B\ddot{\vec{r}}_B=\vec{F}_{AB},
\end{equation}
$\vec{F}_{AB}=-\vec{F}_{BA}=-k^2m_Am_B\vec{r}_{AB}/r_{AB}^3$, and $\vec{r}_{AB}=\vec{r}_B-\vec{r}_A$ it follows that $\ddot{\vec{r}}_{cm}=\vec{0}$. Hence, the center of mass uniformly follows a straight-path. Without loss of generality, $\vec{r}_{cm}=\vec{0}$, and the effective one-body motion reads
\begin{equation}
  \ddot{\vec{r}}_{AB} = -\frac{k^2(m_A+m_B)}{r_{AB}^3}\vec{r}_{AB}.
\end{equation}
The resulting orbital motion is well-known and is given by $r_{AB}(\varphi)=a_{AB}(1-e_{AB}^2)/(1+e_{AB}\cos\varphi)$ with major axis $a_{AB}$, eccentricity $e_{AB}$, and true anomaly $\varphi$.

\begin{figure}[ht]
  \centering\includegraphics[width=0.245\textwidth]{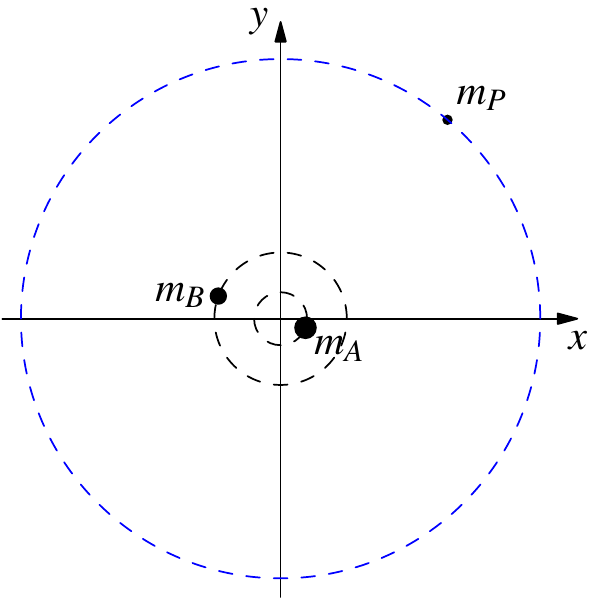}\centering\includegraphics[width=0.245\textwidth]{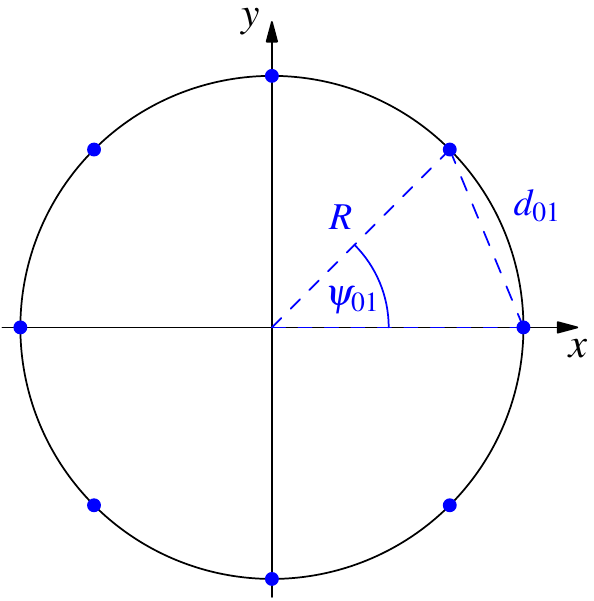}
  \caption{{\it Left:} Planet $P$ orbiting a binary system $(A,B)$. {\it Right:} Particles of equal mass $M$ that are uniformly distributed on a circle of radius $R$ keep on this orbit due to their mutual gravitational attraction.}
  \label{fig:particlesOnRing}
\end{figure}

In case of circular motion, the velocities of the two stars follow from the balance between gravitational attraction and centrifugal force. Thus,
\begin{equation}
v_A^2 = \frac{k^2m_B^2}{(m_A+m_B)r_{AB}},\quad v_B^2 = \frac{k^2m_A^2}{(m_A+m_B)r_{AB}}.
\end{equation}
Similar, the velocity of the planet can be approximated by $v_P^2=k^2(m_A+m_B)/r_P$ orbiting the center of mass of $A$ and $B$.
As an example, the parameters from Doyle et al.~\cite{doyle2011} could be used for the circular motion:
\begin{equation}
  r_{AB}=a_{AB}=0.2243,\quad m_A=0.6897,\quad m_B=0.2026.
\end{equation} 
The planet is described by $m_P=0.333~M_{Jup}/M_{sol}=0.318\cdot 10^{-3}$ and $r_P=0.7048$.

The question that now could arise is: what happens if the distance $r_P$ to the center of mass is reduced?
When will the planet be ejected from the planetary system?

\subsection{Gravitational choreography}
A system of $N$ particles of the same mass $M$ following the same closed trajectory due to their mutual gravitational attraction is called a gravitational choreograph. In case of a circular trajectory of radius $R$, the necessary Keplerian circular velocity of each particle can be determined in the following way.

The gravitational force $\vec{F}_i$ on particle $i$ exerted by all the other particles is given by
\begin{equation}
    \vec{F}_i = \sum\limits_{j=0, j\neq i}^{N-1}\frac{GM^2}{d_{ij}^2}\frac{\vec{r}_{ij}}{|\vec{r}_{ij}|},\quad i=\{0,\ldots,N-1\},
\end{equation}
where $\vec{r}_{ij}=\vec{r}_j-\vec{r}_i$, $|\vec{r}_{ij}|=d_{ij}=2R^2(1-\cos\psi_{ij})$, $\psi_{ij}=2\pi(j-i)/N$, and $\vec{r}_i=R\cos(2\pi i/N)$. Without loss of generality, it suffices to consider $\vec{F}_1$. Because $\vec{F}_1$ must balance the centrifugal force, the Keplerian  circular velocity $v$ then follows from
\begin{equation}
  v^2 = \frac{k^2}{2^{3/2}R}\sum\limits_{j=1}^{N-1}\frac{1}{\sqrt{1-\cos(2\pi j/N)}}.
\end{equation}
Table~\ref{tab:velChoreo} lists the first few (scaled) absolute values of velocity of the particles depending on their number $N$. As can be seen, the velocity is higher the more particles are considered. Numerically, the particles follow the circle only for a limited number of orbits because of the instability of this configuration.
\begin{table}[ht]
  \centering
  \caption{Scaled absolute values of the velocities depending on the number of planets $N$.}
  \label{tab:velChoreo}
  \begin{tabular}{ll@{\hspace*{3em}}ll} \toprule
     $N$ & $v^2\cdot R\cdot 2^{3/2}/k^2$ &  $N$ & $v^2\cdot R\cdot 2^{3/2}/k^2$\\ \hline
     $2$ & $0.707107$ & $6$ & $5.168527$ \\
     $3$ & $1.632993$ & $7$ & $6.518859$ \\
     $4$ & $2.707107$ & $8$ & $7.933369$ \\
     $5$ & $3.892996$ & $9$ & $9.404033$
  \end{tabular}  
\end{table}

Further examples of even more complicated choreographies, also on different trajectories, can be found, for example, in Montgomery~\cite{montgomery2001}, Simo~\cite{simo2001}, Vanderbei~\cite{vanderbei2004}, or \v{S}uvakov and Dmitra\v{s}inovi{\'c}~\cite{suvakov2013}.

\subsection{Planetary ring}
A planetary ring, like the one around Saturn, consists of a huge amount of individual particles of different size and different mass. Through their gravitational interactions clumpy structures might arise. More massive objects like moons can also cause wavy structures due to resonance phenomena as discovered by the Cassini-Huygens mission~\cite{cassini}.

The straightforward numerical simulation of the planetary ring can be done by means of the \emph{DirectNbody} model. To determine the initial velocities of the particles, not only the gravitational potential $\phi_{cm}$ of the central mass $M$ but also the potential of the ring $\phi_{ring}$ has to be taken into account. Thus, at position $\vec{r}'$, the total gravitational potential reads
\begin{equation}
  \phi = \phi_{cm} + \phi_{ring} = -\frac{GM}{|\vec{r}'|} - \iint\limits_{ring}\frac{G\,\diff m}{|\vec{r}'-\vec{r}|},
  \label{eq:ringPoti}
\end{equation}
where $\diff m=\rho r\diff r\diff\varphi$ and $\rho=M_{ring}/[\pi(R_2^2-R_1^2)]$ is the constant surface density. (Details can be found in \ref{appsec:ringVelocities}.) In case of Saturn, for example, the relation between the ring mass and the mass of Saturn is about $M_{Sring}/M_{Sat}\approx 52.8\cdot 10^{-9}$. Hence, the ring potential can be neglected in general.
Nonetheless, it would be interesting to play with different fractions of ring masses. Neglecting the ring potential, a particle's tangential velocity for a circular orbit equals the Keplerian circular velocity: $v^2=k^2/r$ with $k^2=GM$.
\begin{figure}[ht]
  \centering\includegraphics[width=0.35\textwidth]{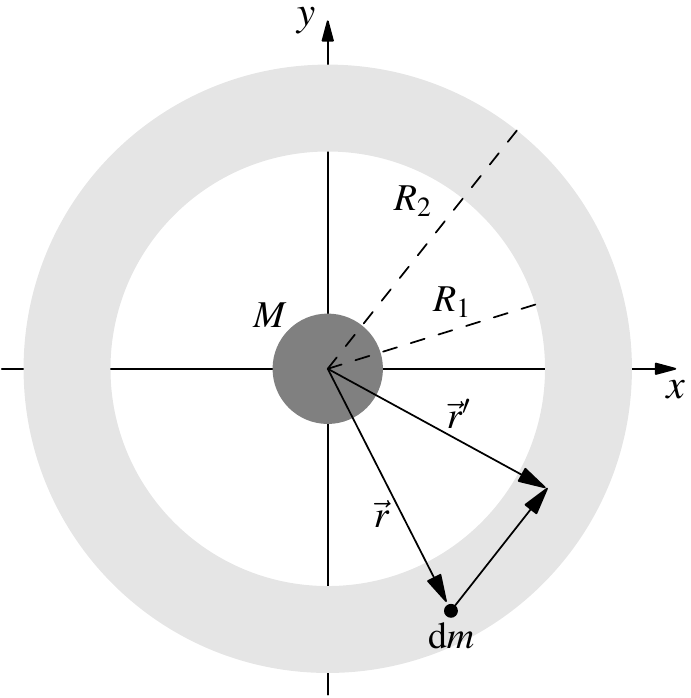}
  \caption{A ring of particles with inner and outer radii, $R_1$ and $R_2$, and surface density $\rho$ around a planet of mass $M$.}
  \label{fig:ringAroundPlanet}
\end{figure}

As an example, consider a planet of unit mass, $M=1$, and $N=3072$ particles with each having $m\approx10^{-8}M$. Then, the particles orbit the planet on nearly circular orbits and the mutual gravitational attractions lead only to small perturbations. But, if the masses of the particles are increased by a factor of $500$, the perturbations become very strong. Increasing the masses by an additional factor $2$ yields to some first clumpy structures of ring particles.
 
Because of the high number of particles, CPU integration of the particles' trajectories becomes extremely slow which makes it necessary to switch to OpenCL-based integration.

\subsection{Double pendulum}
The double pendulum in this example consists of a fixed particle with mass $m_0$ at the ceiling $(x=0, z=l)$, and two particles with masses $m_1$ and $m_2$ which are initially at rest at positions $\vec{x}_1=(0,0)$ and $\vec{x}_2=(0,-l)$. The spring connecting `0' and `1' has spring constant $D_1$ and the one connecting `1' and `2' has spring constant $D_2$. Both springs have rest length $l$.

\begin{figure}[ht]
  \centering\includegraphics[scale=0.8]{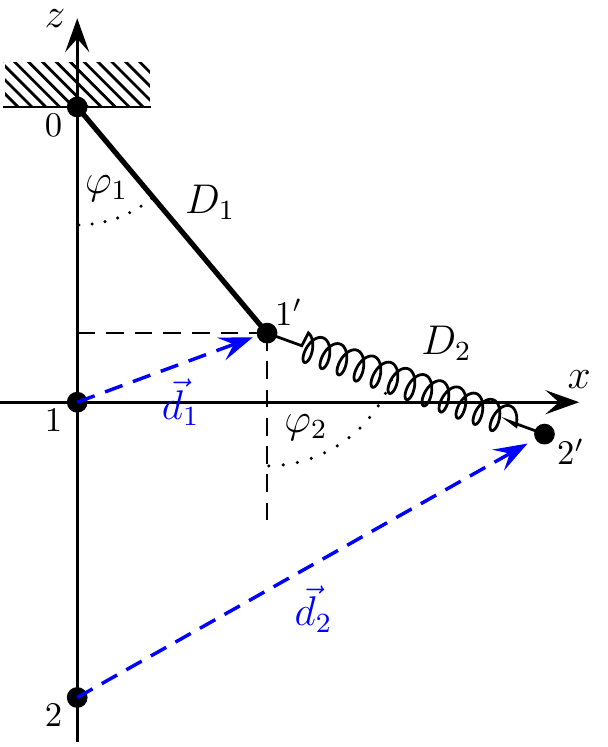}
  \caption{The double pendulum consists of three particles: `0' is fixed, whereas `1' and `2' can move in the $xz$-plane.}
  \label{fig:doublePendulum}
\end{figure}

At the beginning of the simulation, particles `1' and `2' are displaced as shown by the dashed arrows in Fig.~\ref{fig:doublePendulum} with distances $|\vec{x}_{1'}-\vec{x}_0|=|\vec{x}_{2'}-\vec{x}_{1'}|=l$,
\begin{equation}
  \vec{x}_{1'}=\vecTwo{l\sin\varphi_1}{l-l\cos\varphi_1},\quad \vec{x}_{2'} = \vecTwo{l\sin\varphi_1+l\sin\varphi_2}{l-l\cos\varphi_1-l\cos\varphi_2}
\end{equation}
The initial velocities are set to zero, and the driving force is gravitation in the negative $z$-direction.

As an example, $D_1=5\cdot 10^3$, $D_2=5$, $l=1$, $\varphi_1=\varphi_2=0.6$, and the gravitational force is only in the negative $z$-direction, $g=-1.8$. Because of the stiffness of the problem, the step size of the simulation should be small, e.g. $\Delta t\approx 10^{-3}$. If the step size is too high, the simulation crashes and must be restarted. Please note that there is no step size control implemented in the integrators, because this would make the visualization bumpy. And smoothing the particle motion for irregular time steps would be quite expensive.

\subsection{2D membrane}
The eigenmodes of a vibrating rectangular membrane can be determined by solving the two-dimensional wave equation 
\begin{equation}
\frac{1}{c^2}\frac{\partial^2\psi}{\partial t^2}-\frac{\partial^2\psi}{\partial x^2}-\frac{\partial^2\psi}{\partial y^2}=0,\quad\psi\in\mathbb{R}\times\Omega,\,\,\Omega\subset\mathbb{R}^2.
\end{equation}
The product ansatz $\psi(t,x,y)=v(t)\cdot u(x,y)$ yields the two differential equations $\partial_t^2v+\lambda c^2v=0$ and $\Delta u+\lambda u=0$ with a positive constant $\lambda$ and $\Delta=\partial_x^2+\partial_y^2$ being the Laplacian operator in two dimensions. If the membrane is fixed at its boundaries, $\psi(t,0,y)=\psi(t,a,y)=\psi(t,x,0)=\psi(t,x,b)$ with $x\in[0,a]$ and $y\in[0,b]$, the ansatz function has the form
\begin{equation}
  \psi(t,x,y) = \alpha\cos(\omega t+\varphi)\sin(kx)\sin(\hat{k}y)
  \label{eq:membrEM}
\end{equation} 
with $k=r\pi/a$, $\hat{k}=s\pi/b$, for $r,s=\{1,2,\ldots\}$, and some constants $\alpha$ and $\varphi$.

\begin{figure}[ht]
  \centering\includegraphics[width=0.45\textwidth]{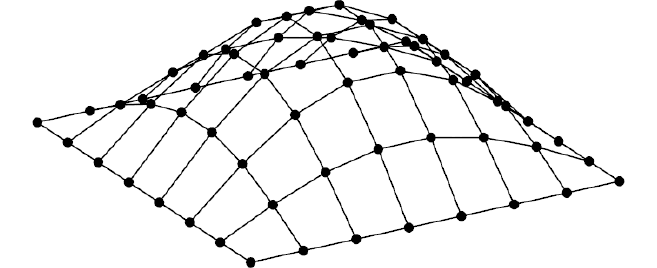}
  \caption{A 2D rectangular membrane with the boundary points being fixed. The interior points, however, can move in all three directions or can be restricted to move only along the z-axis, for example.}
  \label{fig:membrane}
\end{figure}

The rectangular membrane can be approximated by a grid of particles that are connected by springs as shown in Fig.~\ref{fig:membrane}. While the boundary particles are fixed, the motion of the interior particles can be restricted along the $z$-direction or can be unrestricted. From the standard configuration, where all particles are located at $z=0$ and the springs are at their rest lengths, the interior particles are displaced according to the eigenmodes of Eq.~(\ref{eq:membrEM}) in the $z$-direction. However, by means of this na{\"i}ve approach the eigenmodes cannot be reproduced to an acceptable extend, and the grid starts to bounce irregularly already after a few oscillations.

Instead of displacing the particles manually, one could also set a gravitational force to each individual particle and let them move under strong artificial friction until they reach their maximum displacement. Then, friction and force can be set to zero again and the membrane particles oscillate only due to their mutual spring-connections.


\subsection{Cantilever -- 2D beam}
\begin{figure}[ht]
  \centering \includegraphics[width=0.43\textwidth]{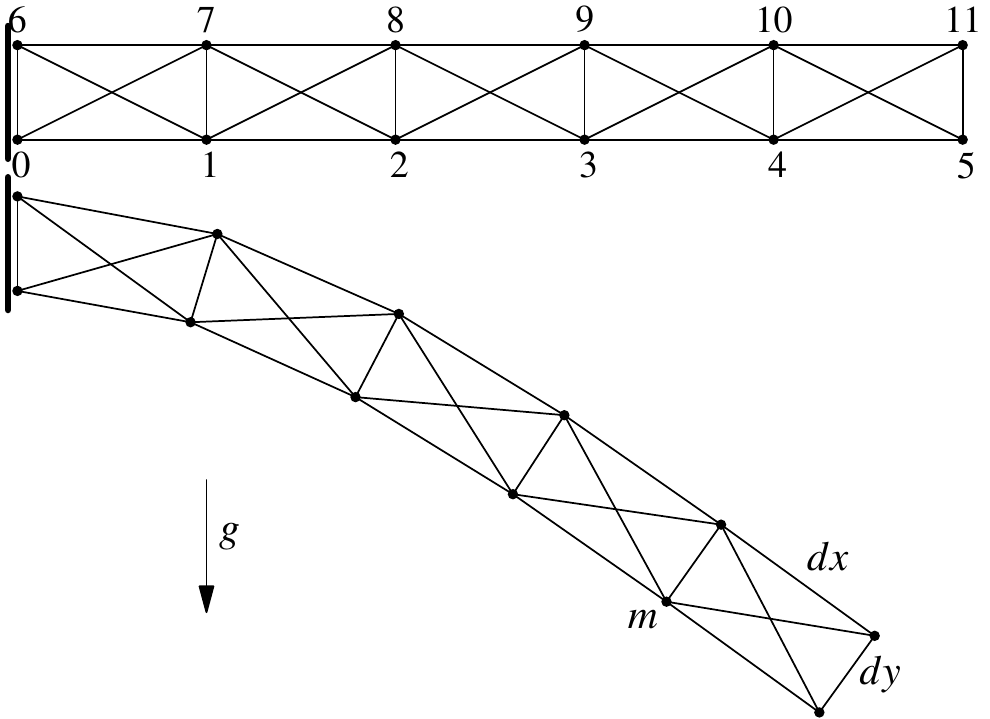}
  \caption{2D cantilever at beginning of the simulation (upper image) and at the lower ``turning'' point (lower image).}
  \label{fig:2dbeam}
\end{figure}

Similar to a 2D membrane, a 2D cantilever can be constructed out of point particles and massless connecting springs. Besides the masses and the spring constants, the influence of the spring lengths $dx$ and $dy$, or the way the masses are connected (with or without cross-connections) can be studied in detail for a varying gravitational force pointing downwards. Figure~\ref{fig:2dbeam} shows a cantilever that is fixed at the wall only with the left most particles `0' and `6'. At the beginning of the simulation, the springs have their rest length $l_{ij}=|\vec{x}_i-\vec{x}_j|$.

\subsection{Newton cradle}
The Newton cradle consists of five balls of equal masses, $m_i=1$, $i=\{1,\ldots,5\}$, where ball `1' is pulled away and is let to fall. When it strikes ball `2', the total energy and momentum of `1' is transported nearly instantaneously to ball `5' that swings away and the procedure starts from the beginning in the reverse direction. In the experiment, however, there is some energy dissipation and the balls come to rest after a few iterations.

\begin{figure}[ht]
  \centering\includegraphics[width=0.45\textwidth]{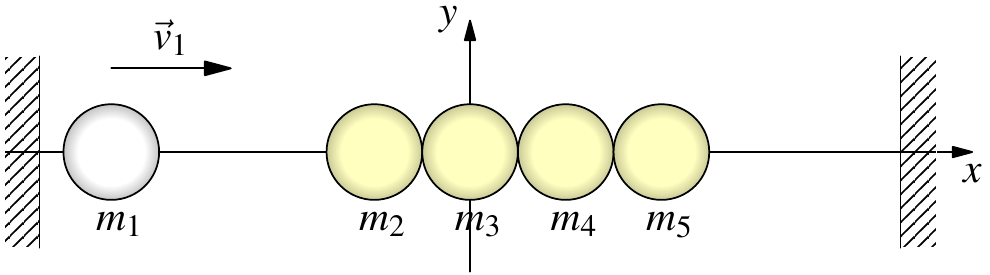}
  \caption{Newton cradle simulation with five balls/spheres.}
  \label{fig:newtonCradle}
\end{figure}

In \emph{MPPhys}, the Newton cradle can be simulated by means of the \emph{HardSpheres} model as shown in Fig.~\ref{fig:newtonCradle}. At the beginning, ball `1' has initial velocity $\vec{v}_1$ and the other balls are at rest. The pendulum effect which let the outer balls return is realized by reflecting walls.

As in the real experiment, the number of balls that are initially pulled away can be changed easily. It is also possible to do a symmetric simulation like, for example, `1' and `5' having velocities $\vec{v}_5=-\vec{v}_1$ and having the same distances to their first-hit balls: $|\vec{x}_1-\vec{x}_2|=|\vec{x}_4-\vec{x}_5|$.

An even more interesting simulation is to use different masses, for example, doubling the mass of particle `1' where the physical explanation becomes non-trivial. See e.g. Kerwin~\cite{kerwin} for a possible explanation.

\subsection{Maxwell-Boltzmann distribution in 2D}
\begin{figure}[ht]
  \includegraphics[width=0.48\textwidth]{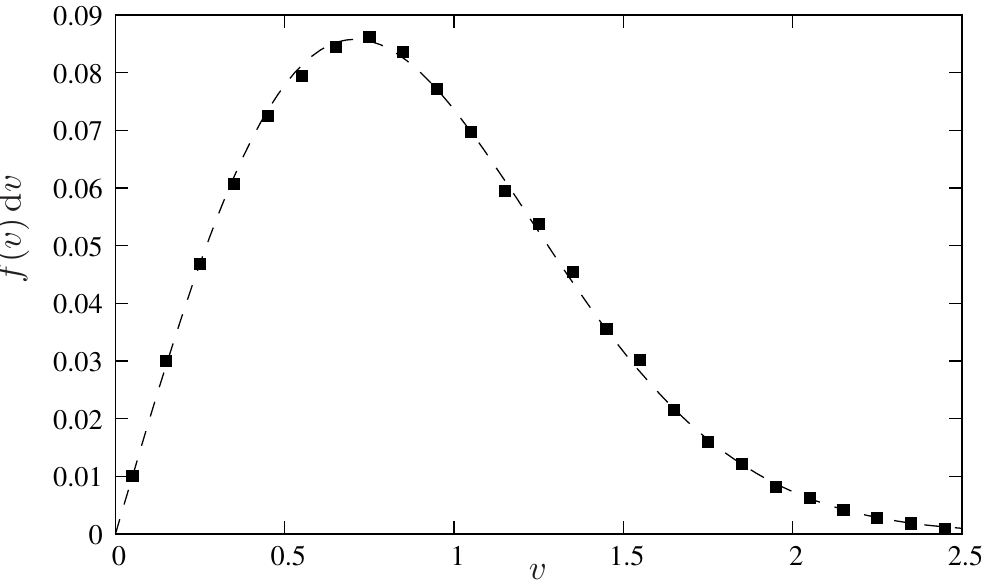}
  \caption{Velocity distribution simulated by the \emph{HardSpheres} model (black dots) at some instance of time during the simulation compared to the Maxwell-Boltzmann distribution in 2D (dashed curve). Here, $b=1$, and the simulation consists of $N=250$ particles with radii $r=0.1$ and initial velocity $v=1$ in a box of size $7.5\times 5.0$.}
  \label{fig:maxBoltz}
\end{figure}

The \emph{HardSpheres} model is also very well suited to simulate an ideal gas. Starting with a random particle distribution where all of the particles have the same velocity magnitude but different directions, the velocity distribution approaches the two-dimensional Maxwell-Boltzmann distribution,
\begin{equation}
  f(v)=\frac{2v}{b^2}\exp\left(-\frac{v^2}{b^2}\right),\quad b^2=\frac{2k_BT}{m},
  \label{eq:maxBoltz}
\end{equation}
already after a few collisions per particle, see Fig.~\ref{fig:maxBoltz}. Here, the Boltzmann constant $k_B$, the temperature $T$, and the particle mass $m$ are chosen such that $b=1$; and the velocity $v$ has the same dimension as $b$.

The same setup can also be used, for example, to study Brownian motion of a massive particle within a `bath' of small particles.

\section{Outlook}
So far, only a simple gravitational N-body simulation, a mass-spring model, and a Discrete Element method based on hard spheres interactions are implemented. But already with these simple simulations, a large number of different particle simulations can be realized as shown in Section~\ref{sec:examples}.

In a future version, several other particle models based on smoothed particle hydrodynamics (SPH), the discrete element method (DEM), molecular dynamics (MD), or the Lattice Boltzmann (LB) method shall be implemented.

\section*{Acknowledgements}
This work was funded by Deutsche Forschungsgemeinschaft (DFG) as part of the Collaborative Research Centre SFB 716.

\appendix

\section{QtMPPhys script engine example}\label{appsec:scriptExp}
The oscillating membrane shown in Fig.~\ref{fig:mpphysGUI} can be generated by the following \emph{QtMPPhys} script. In the first line the particle system and the integrator is set. Then, several parameters are defined for later use. The complete particle data is not set immediately but is stored in the text string {\tt pl} which is handed-over to the system at the end of the script.


{\verbfont
\begin{verbatim}
 // membrane_small.js
 System.Set("MassSpring","LF"); 
 var nx = 8;
 var ny = 8;
 var xStart = -1.0;
 var xEnd   =  1.0;
 var yStart = -1.0;
 var yEnd   =  1.0;
 var mass   =  1.0;
 var radius =  0.01;
 var color  = [1.0,1.0,0.0];
 var dx     = (xEnd-xStart)/(nx-1);
 var dy     = (yEnd-yStart)/(ny-1);
 var pl = new String();
 
 ... (see below)
 
 System.SetData(pl);
\end{verbatim}
}

In the next step, the initial positions as well as the masses, charges, radii, and colors of all particles have to be defined. The {\tt ID} of a particle is used below to set even further parameters. Please note that the {\tt ID} must be in ascending, consecutive order starting from zero! In the \emph{MassSpring} model, the initial position represents the state where the springs have their rest length.
{\verbfont
\begin{verbatim} 
 // pos ID  x y z  m c r  color(r,g,b)
 for(var yy=0; yy<ny; yy++) {
   var y = yStart + yy*dy
   for(var xx=0; xx<nx; xx++) {
     var x = xStart + xx*dx
     pl += sprintf("pos %4d  %12.8f %12.8f %12.8f  
                    %5.3f %5.3f %4.2f  %5.2f %5.2f %5.2f\n",
                    (yy*nx+xx), x,y,0, mass,0,radius, 
                    color[0],color[1],color[2]);
   }
 }  
\end{verbatim}
}

The initial velocity of a particle is zero by default. Here, the velocity is set explicitly for demonstration purpose only.
{\verbfont 
\begin{verbatim}
 // vel ID  vx vy vz
 for(var yy=0; yy<ny; yy++) {
   for(var xx=0; xx<nx; xx++) {
     pl += "vel " + (yy*nx+xx) + " 0.0 0.0 0.0\n";
   }
 }
\end{verbatim} 
}

In the \emph{MassSpring} model, it is necessary that some particles can be fixed such that they can move only in a particular direction or that they are completely static. In this example, the boundary particles are all static and the inner particles can move freely in all directions.
{\verbfont
\begin{verbatim}
 // vfix  ID  fx fy fz
 for(var yy=0; yy<ny; yy++) {
   for(var xx=0; xx<nx; xx++) {
     if (xx==0 || yy==0 || xx==nx-1 || yy==ny-1) {
       pl += "vfix " + (yy*nx+xx) + " 0 0 0\n";
     }
     else {
       pl += "vfix " + (yy*nx+xx) + " 1 1 1\n";
     }
   }
 }
\end{verbatim}
}

The springs connecting the particles have all the same spring constant, frictional coefficient, and color. Hence, only one type of spring is defined. 
{\verbfont
\begin{verbatim}
 // s  ID  D  c  color(r,g,b)
 pl += "s 0 50.0 0.003  0.4 0.4 1.0\n\n";
 for(var yy=0; yy<ny; yy++) {
   for(var xx=0; xx<nx-1; xx++) {
     pl += "sl " + (yy*nx+xx) + " " + (yy*nx+xx+1) + " 0\n";
   }
 }
 for(var xx=0; xx<nx; xx++) {
   for(var yy=0; yy<ny-1; yy++) {
     pl += "sl " + (yy*nx+xx) + " " + ((yy+1)*nx+xx) + " 0\n";
   }
 }
\end{verbatim}
}

The displacements from the initial positions defined above can be used to excite the system from its standard configuration.
{\verbfont
\begin{verbatim} 
 // dis  ID  dx dy dz
 for(var xx=0; xx<nx; xx++) {
   var x = xx/(nx-1.0)*Math.PI;
   for(var yy=0; yy<ny; yy++) {
     var y = yy/(ny-1.0)*Math.PI*1.0;
     var disl = 0.1*Math.pow(Math.sin(x)*Math.sin(y),1.0);
     pl += "dis " + (yy*nx+xx) + " 0 0 " + disl.toFixed(5) + "\n";
   }
 }
\end{verbatim}
}

Finally, some global parameters like a gravitational force, an overall damping constant, or the integration time step {\tt dt} can be defined. 
{\verbfont
\begin{verbatim}
 pl += "grav  0.0 0.0 0.0\n";
 pl += "damp  0.0\n";
 pl += "dt  1e-2";
\end{verbatim}
}

\section{Ring velocities}\label{appsec:ringVelocities}
In polar coordinates, the ring potential $\phi_{ring}$, Eq.~(\ref{eq:ringPoti}), evaluated at position $(r',\varphi')$ reads
\begin{equation}
  \phi_{ring} = -GM\sigma\int_{\varphi=0}^{2\pi}\int_{r=R_1}^{R_2}\frac{r\diff r\diff\varphi}{\sqrt{r^2+r'^2-2rr'(\cos\varphi-\cos\varphi')}},
\end{equation}
where $\rho=M\sigma$ and $\sigma$ being the ratio between the ring mass and the central mass.
Without loss of generality, $\varphi'=0$. Substituting $x=\cos(\varphi/2)$ yields a complete elliptic integral of the first kind,
\begin{equation}
  \phi_{ring} = -4 GM\sigma\int_{r=R_1}^{R_2}\int_{x=0}^1\frac{r\diff r\diff x}{a\sqrt{1-mx^2}\sqrt{1-x^2}}
\end{equation}
with $m=4rr'/a^2$ and $a=r'+r$. Thus, the total gravitational potential at $r'$ reads
\begin{equation}
  \phi(r') = -\frac{GM}{r'} - 4GM\sigma\int_{r=R_1}^{R_2}\frac{r}{r'+r}\mathcal{K}(m)\diff r.
\end{equation}
The initial velocity of a ring particle of mass $\diff{m}$ then follows from the balance between centrifugal and gravitational force,
\begin{equation}
   \frac{\diff{m}\,v^2}{r'} = \diff{m}\,\nabla\phi\quad\text{or}\quad v^2=r'\frac{\partial\phi}{\partial r'}.
\end{equation}

\begin{figure}[ht]
  \includegraphics[width=0.49\textwidth]{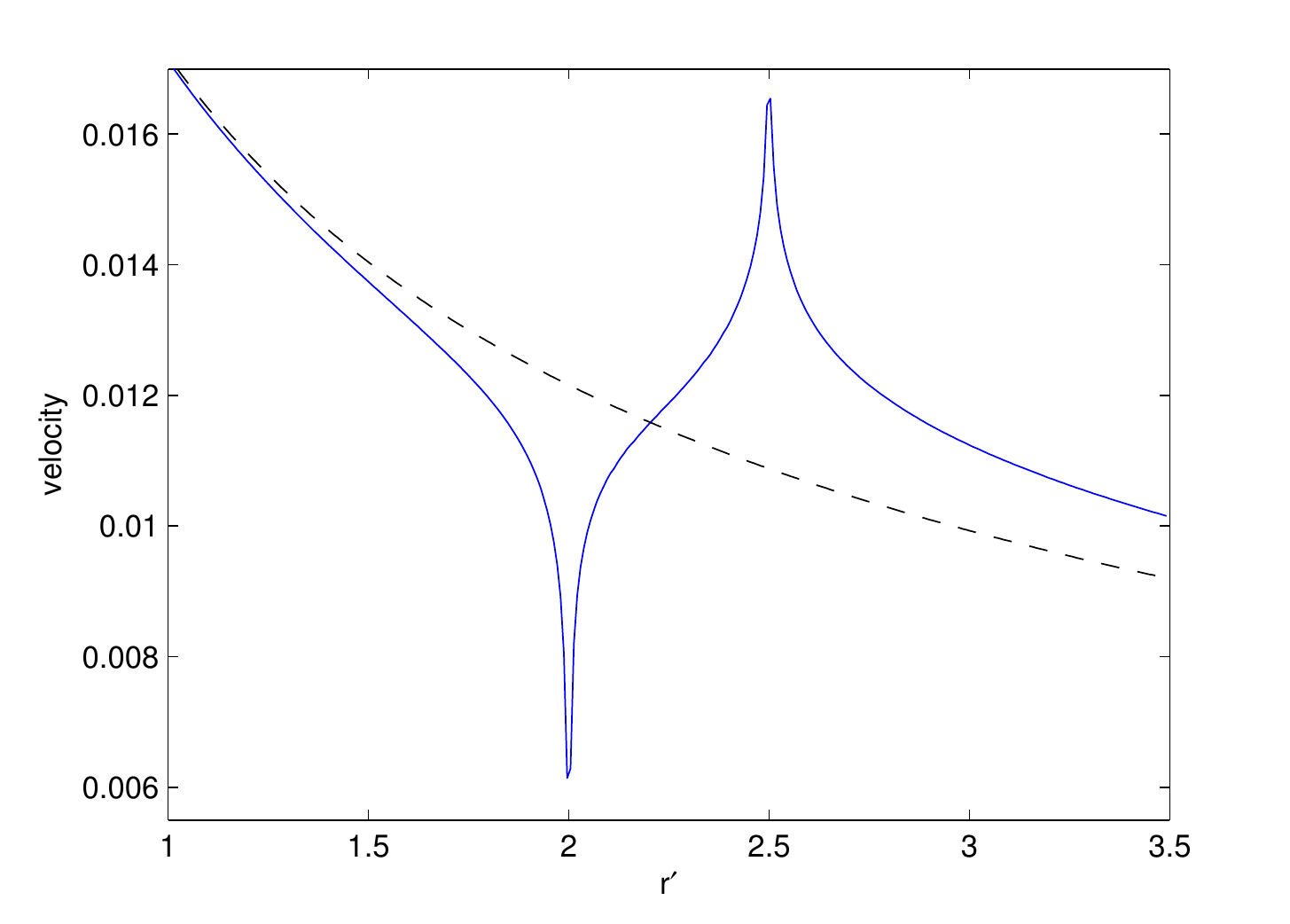}
  \caption{Initial velocity of a particle for a ring with $R_1=2$, $R_2=2.5$, $k=0.0172$. solid line: $\sigma=0.02$, dashed line: $\sigma=0$.}
  \label{fig:ringvel}
\end{figure}

As an example, consider a ring with $R_1=2$, $R_2=2.5$ and a mass ratio $\sigma$ with respect to the central mass $M$. Furthermore, $GM=k^2$ with $k=0.0172$. Then, the initial velocity depending on the distance $r'$ to the center is shown in Fig.~\ref{fig:ringvel}.

Details about elliptic integrals can be found, for example, in Lawden~\cite{lawden} or Armitage and Eberlein~\cite{armitage}.



\end{document}